\documentclass[prb,twocolumn,aps,showpacs]{revtex4}
\usepackage{graphicx}
\usepackage{bm}
\usepackage{amssymb} 
\newcommand{\be}{\begin{equation}}
\newcommand{\ee}{\end{equation}}
\newcommand{\bea}{\begin{eqnarray}}
\newcommand{\eea}{\end{eqnarray}}
\newcommand{\id}{{1}\hspace{-0.3em}\rm{I}}
\newcommand{\si}{\sigma}
\newcommand{\tr}{\tilde{\rho}}
\newcommand{\tx}{\tilde{x}}

\begin{document}

\title{Next-nearest-neighbor spin-spin and chiral-spin
correlation functions in generalized XXX chain}

\author{V. V.  Mkhitaryan and A. G. Sedrakyan}

\affiliation{Yerevan Physics Institute, Br. Alikhanian str.2,
Yerevan 36, Armenia }

\begin{abstract}
We develop a simple technique for calculation of next to nearest
neighbor spin-spin and chiral-spin correlation functions in
inhomogeneous XXX model. Exact expression of the chiral-spin order
parameter as a function of the model parameter, $\omega$, is
analytically found. Using the same method we also calculate the
next to nearest neighbor spin-spin correlation function. In the
limit $\omega\rightarrow 0$ it reproduces the known result for the
vacuum expectation value of the next to nearest neighbor spins in
the standard Heisenberg spin chain. The technique is simple and
can be extended for calculation of next to next to nearest
neighbor correlation functions as well as for calculation of
correlation functions in XXZ model.
\end{abstract}


\maketitle

\section{Introduction}

Presently, calculation of correlation functions in strongly
correlated electron systems is one of the most important tasks in
physics of low dimensions.
Correlation functions of spins at {\em large distances}
are directly related to the observable quantities. They define a
set of critical indices identifying the universality classes of
different phases\cite{Kadanoff-1, Wilson-1, Wilson-2, Kadanoff-2}.
Notably, correlation functions of spins in the homogeneous
Heisenberg spin chain (XXX model) at short distances are not
directly related to the universality class of the phase, but, as
it appeared \cite{MS}, they
play an important role in the perturbative investigations of
ladder models, that are not integrable. An important example is
the Haldane ladder model \cite{haldane}, that can be approached as
an Heisenberg spin chain perturbed with the chiral-spin order
operator, $\vec{\sigma}_{n} (\vec{\sigma}_{n+1} \times
\vec{\sigma}_{n+2})$, (which was considered earlier in connection
with spin-liquid ordered phase in Ref.\onlinecite{wen}) and by the
product of next to nearest neighbor spins,\cite{MS} $\langle
\vec{\sigma}_{n} \vec{\sigma}_{n+2} \rangle$. In order to
investigate the free energy of this long term attractive model and
analyze the phase space,
one needs to know the correlation function of the next to nearest
neighbor spins
and the chiral-spin order parameter $\chi_n=\langle
\vec{\sigma}_{n} (\vec{\sigma}_{n+1} \times \vec{\sigma}_{n+2})
\rangle$.

The aim of the present work is the calculation of abovementioned
correlation functions for a generalized XXX model
\cite{faddeev,ddv,popzvy,zvyagin, frahm,sedrak4} by a new simple
technique. This model is defined and studied in Section II. The
studies of Section II involve analysis of the Hamiltonian(s) and
other conserved currents, construction of the Transfer matrix and
obtaining the set of Bethe equations that describe the energy
spectrum. Two particular cases of the generalized XXX model are
the ladder systems with $N$ sites at each chain given by
Hamiltonian operators

\begin{eqnarray}
\label{zvmer} {\mathcal H}=2\sum\limits_{n=1}^{2N}\left[
\vec{\sigma}_{n} \vec{\sigma}_{n+1}-1 \right ]+
\omega^2\sum\limits_{n=1}^{2N}\left[ \vec{\sigma}_{n}
\vec{\sigma_{n+2}}-1 \right]\nonumber\\
\qquad\qquad+\omega\sum\limits_{n=1}^{2N}(-1)^n
\vec{\sigma}_{n-1}(\vec{\sigma}_{n}\times\vec{\sigma}_{n+1}),
\end{eqnarray}
and
\begin{eqnarray}
\label{mer} {\mathcal
H}^{\prime}=\omega^2\sum\limits_{n=1}^{2N}(-1)^n\left[
\vec{\sigma}_{n}
\vec{\sigma_{n+2}}-1 \right]\qquad\qquad\nonumber\\
\qquad\qquad\quad-\omega\sum\limits_{n=1}^{2N}
\vec{\sigma}_{n-1}(\vec{\sigma}_{n}\times\vec{\sigma}_{n+1}).
\end{eqnarray}
 As a manifestation of the effective workability of the
developed technique in Section III we first calculate the exact
correlation functions of the next to nearest neighbor spins,
$\xi(\omega)=\langle \vec{\sigma}_{n}
\vec{\sigma}_{n+2}-1\rangle$, as a function of $\omega$, which for
the Hamiltonian Eq.~(\ref{zvmer}) has the following asymptotic
behavior:
\begin{eqnarray}
\label{behavior} \xi(\omega)\!=\!\left\{
\begin{array}{l}
-16\log (2)+9\zeta(3)-1,41 \omega^2 ,\nonumber\\
\qquad\qquad\qquad\qquad\qquad\;\;\;\;\;\text{for}\;\omega\ll 1\\
-2,77 + \frac{1,77}{\omega^2}, \nonumber\\
\qquad\qquad\qquad\qquad\qquad\;\;\;\;\;\text{for}\;\omega\gg 1.
\end{array}
\right.\\\!\!\!
\end{eqnarray}
Here $\zeta(x)$ is the  Riemann zeta function. In the limit
$\omega=0$,
we reproduce the known result for the expectation value of next to
nearest neighbor exchange operator in the standard XXX chain,
first obtained by involving the Hubbard model\cite{takahashi77}.

More recently,  within the general approach of multiple integral
representation of correlation functions, that was formulated in
Ref.~\onlinecite{JM} and investigated further in
Refs.~\onlinecite{KMT} and \onlinecite{KMT1}, correlation
functions $\langle \vec{\sigma}_{n} \vec{\sigma}_{n+3} \rangle$
and $\langle \vec{\sigma}_{n} \vec{\sigma}_{n+4}\rangle$ were
evaluated\cite{SSNT,BST}.
These functions were evaluated for $XXX$ model in zero magnetic
field. One of the advantages of our technique is the possibility
to include also external magnetic field. This can be done by
involving the magnetic field into the integral equations for the
densities. In some cases such integral equations can be solved by
approximate methods \cite{JNW}.


Our main result, however, is the analytical calculation of the
chiral-spin order parameter, $\langle \vec{\sigma}_{n}
(\vec{\sigma}_{n+1} \times \vec{\sigma}_{n+2})
\rangle=(-1)^n\chi(\omega)$, as a function of staggering
parameter, $\omega$. The full expression and the derivation of
$\chi_n$ are presented in Section III. Here we present only the
asymptotes of the function $\chi(\omega)$,
\begin{eqnarray}
\label{behavior-1} \chi(\omega)\!=\!\left\{
\begin{array}{l}
\!\bigl[8\gamma+8\psi(1/2)-\psi^{\prime\prime}(1/2)+
\!\psi^{\prime\prime}(1)\bigr]\omega ,\nonumber\\
\qquad\qquad\qquad\qquad\qquad\;\;\;\;\text{for}\;\omega\ll 1,\\
\frac{3,545}{\omega},\;\;\;\;\;\text{for}\;\omega\gg 1,
\end{array}
\right.\\ \!\!\!
\end{eqnarray}
where $\gamma$ is the Euler constant, the function $\psi$ is the
Digamma function and the coefficient of linearity for small
$\omega$ is $\approx 3,33.$

We would like to note that this correlation function was
previously investigated numerically \cite{frahm} and the match
between curves plotted from our exact result (Fig.2) and from the
numerical simulations is perfect.

\section{Inhomogeneous Chain and a Family of Commuting Operators}

In one dimension, there is a variety of quantum exactly solvable
models of interacting spins. Some of these models involve
interactions between nearest neighbor spins and also spins that
are far from each other. Usually the further neighbor interactions
and other non-localities come from additional anisotropy
parameters. We will consider a family of models with nearest
neighbor, next to nearest neighbor and triangular (zig-zag)
interactions, which stem from the Transfer matrix with the shift
of the spectral parameters at each second sites: \bea \label{tm}
\Theta(\lambda;\omega)&=&L_{2N,a}(\lambda)L_{2N-1,a}(\lambda-\omega)\cdots\\
\nonumber &&\cdots L_{2,a}(\lambda)L_{1,a}(\lambda-\omega). \eea
Here $L_{a,b}(\lambda)$ obeys the Rational Yang-Baxter relations
\bea \label{YB}
&&R_{a_1,a_2}(\lambda-\mu)L_{n,a_1}(\lambda)L_{n,a_2}(\mu) \\
\nonumber && \qquad \qquad = L_{n,a_2}(\mu)
L_{n,a_1}(\lambda)R_{a_1,a_2}(\lambda-\mu) \eea
with rational \bea
\label{LR} L_{n,a}(\lambda)&=&(\lambda-i/2)\id_{n,a}+iP_{n,a}, \\
\nonumber R_{a,b}(\lambda) &=&\lambda\id_{a,b}+iP_{a,b}. \eea
The
permutation operator is given in terms of Pauli matrices as
$$P_{a,b}=\frac
12(\id_a\otimes\id_b+\sum_\alpha\si^\alpha\otimes\si^\alpha).
$$ With this construction, one has a commuting one parametrical family of
transfer matrices $\tau(\lambda;\omega)=tr_a\Theta(\lambda;\omega):$
$$[\tau(\lambda;\omega),\tau(\mu;\omega)]=0.$$
This is a well known picture, see e.g., \cite {faddeev,ddv,
zvyagin}, while we want to apply this in a slightly different
contents. In order to be shorter we will miss some proofs and refer
the reader to the nice review by L. D. Faddeev \cite {faddeev}.

The interesting feature of the Transfer matrix
$\Theta(\lambda;\omega)$, given by Eq.~(\ref{tm}), is that instead
of usual XXX Hamiltonian this yields two different local
Hamiltonian operators, $H_1$ and $H_2$, that are proportional to
the logarithmic derivative of $\tau$ at two different points:
$\lambda=i/2$ and $\lambda=\omega+i/2$. Respectively, their
explicit forms are

\bea
\label{Ham1}
H_1&=&2i(1+\omega^2)\partial_\lambda \ln
\tau \big |_{\lambda=i/2}-N(2+\omega^2-2i\omega)\nonumber\\
&=&\sum\limits_{n=1}^{2N}\left[  \vec{\sigma}_{n}
\vec{\sigma}_{n+1}-1 \right ]+
\sum\limits_{k=1}^{N}\left\{\omega^2\left[  \vec{\sigma}_{2k}
\vec{\sigma_{2k+2}}-1 \right] \right.\nonumber \\
&-&\left. \omega\vec{\sigma}_{2k}(\vec{\sigma}_{2k+1}\times\vec{\sigma}_{2k+2})\right\},\\
\nonumber\\
\nonumber \\
H_2&=&2i(1+\omega^2)\partial_\lambda\ln\tau
\big |_{\lambda=i/2+\omega}-N(2+\omega^2+2i\omega)\nonumber\\
&=&\sum\limits_{n=1}^{2N}\left[  \vec{\sigma}_{n}
\vec{\sigma_{n+1}}-1 \right]+
\sum\limits_{k=1}^{N}\left\{\omega^2\left[ \vec{\sigma}_{2k-1}
\vec{\sigma_{2k+1}}-1 \right]\right. \nonumber \\
&+&\left.\omega\,\vec{\sigma}_{2k-1}(\vec{\sigma}_{2k}\times\vec{\sigma}_{2k+1})\right\}.
\eea These operators are commuting as they belong to the same
commuting family. It is straightforward that the Hamiltonian
operators ${\mathcal H}=H_1-H_2$ and ${\mathcal
H}^{\prime}=H_1+H_2$ are exactly diagonalizable in the same
framework. Their explicit forms are given by Eqs. (\ref{zvmer})
and (\ref{mer}).

Let {\em quasi- shift operators} be the monodromy matrices at the
points $\lambda=i/2$ and $\lambda=i/2+\omega$:

\bea \label{shops}
U_+&=& tr_a\Theta(i/2;\omega),\\ \nonumber
U_-&=&
tr_a\Theta(i/2+\omega;\omega).
\eea
They obey the relation
\be
\label{shs} (1+\omega^2)^N V^2=U_+U_-=e^{iP},
\ee
where
\be
\label{V} V=P_{1,2}P_{2,3}\dots P_{2N-1,2N}
\ee
is a shift  on one
unit $n\rightarrow n+1$, and $P$ is the physical momentum, which
governs the shift $n\rightarrow n+2$. Being defined in this way,
the quasi-shift operators commute with the whole family of
transfer matrices $\tau(\mu;\omega)$.

Derivation of the Bethe Ansatz Equations (BAE) for
$\tau(\mu;\omega)$ is standard, see e.g., Ref.\cite{faddeev}.
Starting from the reference state with all $2N$ spins up one can
generate the eigenvectors of the transfer matrix in the sector
with $M$ overturned spins, parametrized by $M$ complex rapidities
$\lambda_n$ which obey BAE \be \label{BAE} \left(\frac
{(\lambda_n+i/2)(\lambda_n-\omega+i/2)}{(\lambda_n-i/2)(\lambda_n-\omega-i/2)}\right)^N=\prod_{k\neq
n}^M \frac {\lambda_n-\lambda_k+i}{\lambda_n-\lambda_k-i}. \ee The
eigenvalues of the transfer-matrix $\tau(\mu;\omega)$ have the
form \bea \label{tme} t(\mu)&=& [(\mu+\frac i2)(\mu-\omega+\frac
i2)]^N\prod_{n=1}^M\frac{\mu-\lambda_n-i}{\mu-\lambda_n} \nonumber \\
&+&
[(\mu-\frac i2) (\mu-\omega-\frac
i2)]^N\prod_{n=1}^M\frac{\mu-\lambda_n+i}{\mu-\lambda_n}.
\nonumber \\
\eea
This
gives the quasiparticle momentum in the form
\be
\label{me}
e^{iP}=\prod_{n=1}^M\frac
{(\lambda_n+i/2)(\lambda_n-\omega+i/2)}{(\lambda_n-i/2)(\lambda_n-\omega-i/2)},
\ee
and eigenvalues of $H_1$ and $H_2$ as follows:
\bea
\label{e1e2}
E_1(\{\lambda\},\omega)&=&-2(1+\omega^2)\sum_n\frac 1
{\lambda_n^2+1/4}, \\
E_2(\{\lambda\},\omega)&=&-2(1+\omega^2)\sum_n\frac 1
{(\lambda_n-\omega)^2+1/4}. \nonumber \eea The corresponding
eigenenergies of Eqs. (\ref{zvmer}) and (\ref{mer}) are $E_1+E_2$
and $E_1-E_2$ respectively. Now the picture is in some sense
complete and one is in position to infer the thermodynamics of
these models, based on the BAE and the energy relations. The
particular Hamiltonian Eq.~(\ref{zvmer}) was introduced in
Ref.~\onlinecite {popzvy} and analyzed in details \cite {zvyagin,
frahm}. It has a singlet ground state with massless excitations.
By involving a magnetic field with the Zeeman coupling, the system
undergoes two phase transitions; two critical phases with
different universality classes are discussed in
Ref.~\onlinecite{frahm1}. The XXZ generalization of the model
(\ref{mer}) was defined and investigated in \cite {sedrak4,
mksed}.

\section{Chiral-spin and other correlation functions}
 It is well
known that the expectation values of operators not commuting with
the Hamiltonian are not easily accessible within the framework of
Bethe Ansatz. In the case of inhomogeneous chain under
consideration we have the additional parameter $\omega$, which
breaks the translational invariance $n\rightarrow n+1$ and gives a
possibility to calculate some simplest expectation values, that
are valid also in the limit $\omega\rightarrow 0$, corresponding
to the well known XXX case. For our purposes we need the
eigenvalues of quasi-translation operators, which follow from
(\ref{tme}) and the definition (\ref{shops}): \bea \label{uu}
u_+&=&(-1)^N(1+i\omega)^N\prod_{n}\frac
{\lambda_n+i/2}{\lambda_n-i/2}, \\
u_-&=&(-1)^N(1-i\omega)^N\prod_{n}\frac
{\lambda_n-\omega+i/2}{\lambda_n-\omega-i/2}. \nonumber
\eea

One can differentiate Eq.~(\ref{shops}) with respect to $\omega$
and get the relations \bea \label{o1}
\partial_\omega U_+&=&\sum_{k=1}^N \frac
{iP_{2k,2k-1}+\omega}{1+\omega^2}U_+ \nonumber \\
&=&U_+\sum_{k=1}^N \frac {iP_{2k,2k+1}+\omega}{1+\omega^2},
\eea
\bea
\label{uu1}
\partial_\omega U_-&=&-\sum_{k=1}^N  \frac
{iP_{2k,2k+1}-\omega}{1+\omega^2}U_- \nonumber \\
&=&-U_-\sum_{k=1}^N \frac {iP_{2k,2k-1}-\omega}{1+\omega^2}
\eea

Let us use the relation, that is always valid when one has a
parameter-dependent operator $\hat{O}(\eta)$ with the spectrum
$o_n(\eta)$ and normalized eigenstates $|n\rangle$ \bea \label{do}
\langle n|\partial_\eta\hat{O}(\eta)|n\rangle=\partial_\eta
o_n(\eta). \eea

Then we find that
\bea
\label{1cor}
&&\langle \{\mu\}|\vec{\sigma}_{n}
\vec{\sigma}_{n+1}|\{\mu\}\rangle \nonumber \\
&=&\mp \frac 2N i(1+\omega^2)
\partial_\omega\ln[u_{\pm}(\omega)] +2i\omega -1 \nonumber \\
&=& 1 - 2\frac{(1+\omega^2)}N\sum_{m=1}^M
\frac{\partial_\omega\mu_m}{\mu^2_m+1/4}, \eea where $\{\mu\}$ is
any set of BAE solution. Given in the above form, it is readily
calculable. It does not depend on $n$, even on the parity of $n$
and shows that the model Eq.~(\ref{zvmer}) can't have a dimerized
phase.

The next expectation value that we are going to evaluate is the
next to nearest neighbor (NNN) exchange, $\langle
\{\mu\}|\vec{\sigma}_{n} \vec{\sigma}_{n+2}|\{\mu\}\rangle$. For
this purpose we divide the two Hamiltonian operators
Eq.~(\ref{Ham1}) by $\omega$ and apply Eq.~(\ref{do}). Subtracting
contributions of nearest neighbor terms with the use of
Eq.~(\ref{1cor}) we obtain for even or odd site numbers the
following relations: \vspace{3mm}

\begin{widetext}
\bea
\label{2cor}
\langle
\{\mu\}|\vec{\sigma}_{2k} \vec{\sigma}_{2k+2}-1|\{\mu\}\rangle
&=&
\frac 1N\partial_\omega\frac{E_1(\{\mu\},\omega)}{\omega} +\frac
2{\omega^2}\langle \{\mu\}|\vec{\sigma}_{n}
\vec{\sigma}_{n+1}-1|\{\mu\}\rangle\\
&=& -\frac
4N\sum_{m=1}^M\left\{
\frac{1+\omega^2}{\omega^2}\frac{\partial_\omega\mu_m}{\mu^2_m+\frac
14} +\frac{\omega^2-1}{2\omega^2}\frac{1}{\mu^2_m+\frac 14}
-\frac{1+\omega^2}{\omega}\frac{\mu_m\partial_\omega\mu_m}{(\mu^2_m+\frac
14)^2} \right\}\nonumber
\eea
and
\bea
\label{2cor1}
\langle \{\mu\}|\vec{\sigma}_{2k-1}
\vec{\sigma}_{2k+1}&-&1|\{\mu\}\rangle = \frac
1N\partial_\omega\frac{E_2(\{\mu\},\omega)}{\omega} +\frac
2{\omega^2}\langle \{\mu\}|\vec{\sigma}_{n}
\vec{\sigma}_{n+1}-1|\{\mu\}\rangle\\ &=& -\frac
4N\sum_{m=1}^M\left\{
\frac{1+\omega^2}{\omega^2}\frac{\partial_\omega\mu_m}{\mu^2_m+\frac
14} +\frac{\omega^2-1}{2\omega^2}\frac{1}{(\mu_m-\omega)^2+\frac 14}
-\frac{1+\omega^2}{\omega}\frac{(\mu_m-\omega)(\partial_\omega\mu_m-1)}{((\mu_m-\omega)^2+\frac
14)^2} \right\}.\nonumber \eea
\end{widetext}
It is yet unknown how to
perform such summations in the case of finite $N$ and $M$
analytically. Instead we can evaluate the sums in the important
case of thermodynamic limit, \be \label{tl}
N\rightarrow\infty,\quad M\rightarrow\infty,\quad\frac MN=const,
\ee when solutions of Eq.~(\ref{BAE}) form bound states called
strings \cite{takahashi}. In our rational case strings with
arbitrary length $n$ are possible. Consider the case where one has
$M_n$ bound states of $n$-strings \bea \label{string}
\lambda_\alpha^{n,j}&=&\lambda_\alpha^{n}+\frac
i2(n+1-2j)+O(\exp(-|\delta| N)),\nonumber  \\
\alpha&=&1,..,M_n,\qquad j=1,..,n, \eea with real parameters of
centers $\lambda_\alpha^{n}$. Then one can take the product of $n$
BAE for the same $n$-string and obtain an equation for real
centers. The logarithm of these equations gives \bea
\label{logBAE}
&&\theta(\lambda_\alpha^n/n)+\theta((\lambda_\alpha^n-\omega)/n)\\
&=&\frac{2 \pi}{N} I_\alpha^n+\frac{1}{N}\sum_{(m,\beta)\ne
(n,\alpha)}\Theta_{nm}(\lambda_\alpha^n-\lambda_\beta^m), \nonumber
\eea
where
$\theta(\lambda)\equiv 2\tan^{-1}[2\lambda]$
and
\begin{widetext}
\bea
\label{theta}
\Theta_{nm}(\lambda)\equiv\left\{\matrix{\theta({\lambda\over \vert
n-m\vert}) +2\theta({\lambda\over \vert n-m \vert
+2})+...+2\theta({\lambda\over n+m-2}) +\theta({\lambda\over
n+m}) \qquad \qquad {\rm for}\quad n\ne m,\cr \nonumber \\
2\theta({\lambda\over
2})+2\theta({\lambda\over 4})+...+2\theta({\lambda\over 2n-2})
+\theta({\lambda\over 2n})\qquad \qquad \qquad \qquad \qquad {\rm for}\quad  n=m.}\right.
\eea
\end{widetext}
$I_\alpha^n$ is an integer (half-odd integer) if $N-M_n$ is odd
(even) and satisfies
\bea
\label{vv}
\vert I_\alpha^n\vert &\leq & {1\over 2}(N-1-\sum_{m=1}^\infty t_{nm}M_m),\nonumber \\
t_{nm} & \equiv & 2 Min(n,m)-\delta_{nm}. \eea

In the thermodynamic limit (\ref{tl}), it is convenient to define
distribution functions of $n$-strings $\rho_n(\lambda)$ and holes of
$n$-string $\tr_n(\lambda)$; the number of strings and holes between
$\lambda$ and $\lambda+d\lambda$ is $\rho_n(\lambda)Nd\lambda$ and
$\tr_n(\lambda)Nd\lambda$, respectively.  From Eqs.(\ref{logBAE})
one obtains a system of integral equations \bea \label{iero}
a_n(\lambda)&+& a_n(\lambda-\omega)
=\rho_n(\lambda)+\tr_n(\lambda)\\
&+&\sum_m\int^\infty_{-\infty}
T_{nm}(\lambda-\mu)\rho_m(\mu)d\mu,\nonumber
\eea
where $T_{nm}(\lambda)$ is a
function defined by
\begin{widetext}
\bea
\label{TT}
T_{nm}(\lambda)\equiv
\left\{\matrix{a_{|n-m|}(\lambda)+2a_{|n-m|+2}(\lambda)
+2a_{|n-m|+4}(\lambda)+...\cr \nonumber\\
+2a_{n+m-2}(\lambda)+a_{n+m}(\lambda)~~~~ \qquad  \qquad  \qquad  \qquad
{\rm for}\qquad  n\ne m,\cr \nonumber\\
2a_{2}(\lambda)+2a_{4}(\lambda)+...
+2a_{2n-2}(\lambda)+a_{2n}(\lambda)\qquad {\rm for}\qquad n=m.}\right.
\eea
\end{widetext}
and $a_n(\lambda)$ is a function defined by
$$
a_n(\lambda)\equiv{1\over \pi}{2n\over 4\lambda^2+n^2}.
$$

In order to describe $\omega$- derivatives of $\lambda_\alpha^n$
in the thermodynamic limit, we introduce a new function, $F_n$,
as
\bea \label{newfunc}
\lim_{N\rightarrow\infty}\partial_{\omega}\lambda_\alpha^nF_n(\lambda,\omega). \eea For briefness we will miss the explicit
$\omega$- dependence of $F$. In order to find a characteristic
integral equation for this function, one can differentiate
(\ref{logBAE}) with respect to $\omega$ and use (\ref{iero}). In
this way one finds:
\bea \label{ienf}
a_n(\lambda-\omega)&=&F_n(\lambda)[\rho_n(\lambda)+\tr_n(\lambda)]\\ \nonumber\\
&+&\sum_m\int^\infty_{-\infty}
T_{nm}(\lambda-\mu)F_m(\mu)\rho_m(\mu)d\mu.\nonumber
\eea
Now we can rewrite
the R.H.S. sums in (\ref{2cor}) in terms of integrals:
\begin{widetext}
\bea
\label{2corint}
\langle \{\mu\}|\vec{\sigma}_{2k}
\vec{\sigma}_{2k+2}&-&1|\{\mu\}\rangle \\
&=&
-8\pi\sum_{n=1}^\infty\int\rho_n(\mu)d\mu\left\{
\frac{1+\omega^2}{\omega^2}a_n(\mu)F_n(\mu)
+\frac{\omega^2-1}{2\omega^2}a_n(\mu)+\frac{1+\omega^2}{2\omega}a_n^\prime(\mu)F_n(\mu)
\right\},\nonumber
\eea
and
\bea
\label{2corint1}
&&\langle \{\mu\}|\vec{\sigma}_{2k-1}
\vec{\sigma}_{2k+1}-1|\{\mu\}\rangle\\
&=&
-8\pi\sum_{n=1}^\infty\int\rho_n(\mu)d\mu\left\{
\frac{1+\omega^2}{\omega^2}a_n(\mu)F_n(\mu)
+\frac{\omega^2-1}{2\omega^2}a_n(\mu-\omega)+
\frac{1+\omega^2}{2\omega}a_n^\prime(\mu-\omega)(F_n(\mu)-1)
\right\}.\nonumber
\eea
\end{widetext}
Evaluation of the expectation value of triple interaction terms,
the chiral-spin order parameter, can be done in a similar way. The
answer is \bea \label{3corint} &&\langle
\{\mu\}|\vec{\sigma}_{2k}(\vec{\sigma}_{2k+1}\times\vec{\sigma}_{2k+2})|\{\mu\}\rangle\nonumber\\
&=& -4\pi\sum_{n=1}^\infty\int\rho_n(\mu)d\mu\left\{
4\frac{1+\omega^2}{\omega}a_n(\mu)F_n(\mu)\right. \nonumber \\
&-&\left.\frac
2{\omega}a_n(\mu)+(1+\omega^2)a_n^\prime(\mu)F_n(\mu)
\right\},
\eea
and
\bea
\label{3corint1}
&&\langle
\{\mu\}|\vec{\sigma}_{2k-1}(\vec{\sigma}_{2k}\times\vec{\sigma}_{2k+1}|\{\mu\}\rangle\\
&=& 4\pi\sum_{n=1}^\infty\int\rho_n(\mu)d\mu\left\{
4\frac{1+\omega^2}{\omega}a_n(\mu)F_n(\mu)\right.\nonumber\\
&-&\left.\frac 2{\omega}a_n(\mu-\omega)+
(1+\omega^2)a_n^\prime(\mu-\omega)(F_n(\mu)-1) \right\}.\nonumber
\eea

\begin{figure}[t]
\centerline{\includegraphics[width=55mm,angle=0,clip]{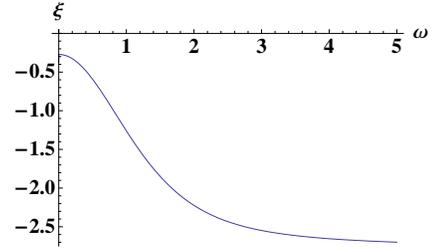}}
\caption{Next to nearest neighbor correlation function
$\xi(\omega)$ versus inhomogeneity parameter $\omega$.}
\end{figure}

For definiteness, let us evaluate the NNN expectation value for
the ground state of the Hamiltonian Eq.~(\ref{zvmer}). For this
state, the densities are found to be zero for all the $n$- strings
with $n=2,3,..$ and $1$-holes \cite{frahm}. The system
Eq.~(\ref{iero}) reduces to the following simple integral equation
for $n=1$: \be \label{iero1}
a_1(\lambda)+a_1(\lambda-\omega)=\rho_1(\lambda)+\int^\infty_{-\infty}
T_{11}(\lambda-\mu)\rho_1(\mu)d\mu. \ee Its solution \be
\label{vacrho} \rho_1(\lambda)=\frac 1{2\cosh\pi\lambda}+\frac
1{2\cosh\pi(\lambda-\omega)}, \ee and the corresponding solution
to (\ref{ienf}) \bea \label{vacf}
F_1(\lambda)\rho_1(\lambda)=\frac 1{2\cosh\pi(\lambda-\omega)}
\eea can be found by the Fourier transform. The integrals in
(\ref{2corint}) can be easily transformed to the following
expression for the NNN expectation value: \bea \label{nnn}
&&\xi(\omega)=\langle \{\mu\}|\vec{\sigma}_{n}
\vec{\sigma}_{n+2}-1|\{\mu\}\rangle\\
&=&-\frac
{3\omega^2+1}{\omega^2}I(\omega)
-\frac {\omega^2-1}{\omega^2}I(0)
-\frac {\omega^2+1}{\omega}\partial_\omega I(\omega), \nonumber
\eea
 where
$I(\omega)$ is the integral which can be expressed via Digamma
functions, $\psi$, as \bea \label{dg} I(\omega)&=&
\int^\infty_{-\infty}\frac{dx}{(x^2+1/4)\cosh\pi(x-\omega)}\nonumber \\
&=&\psi(1+i\frac\omega 2)
-\psi(\frac 12+i\frac\omega 2)  \nonumber \\
&+& \psi(1-i\frac\omega 2) -\psi(\frac
12-i\frac\omega 2) .
\eea

We see that though the translational
invariance $n\rightarrow n+1$ is broken,
the vacuum expectation
value of the NNN exchange operator does not depend on the site
parity. Up to the sign factor, the same is valid for the triple
interaction terms: $ \langle \vec{\sigma}_{2k}
(\vec{\sigma}_{2k+1} \times\vec {\sigma}_{2k+2}\rangle=- \langle
\vec{\sigma}_{2k-1} (\vec{\sigma}_{2k} \times\vec
{\sigma}_{2k+1}\rangle$.

In the limit $\omega\rightarrow 0$, from Eq.~(\ref{nnn}), one will
recover the known result for the expectation value of NNN exchange
operator of Heisenberg XXX isotropic chain, \be \label{tr} \langle
\vec{\sigma}_{n} \vec{\sigma}_{n+2}\rangle=1-16\log (2)+9\zeta(3).
\ee This was calculated from the ground state energy of the
Hubbard model in Ref.~\onlinecite{takahashi77}. We present the
function $\xi(\omega)$ in Fig.1.

The chiral-spin order parameter $\langle \{\mu\}| \vec{\sigma}_{n}
(\vec{\sigma}_{n+1}\times\vec{\sigma}_{n+2})|\{\mu\}\rangle$ is
also important, as it defines the measure of chirality of the
state. Substituting the densities $\rho$ and $F$ for the ground
state, Eqs. (\ref{vacrho}) and (\ref{vacf}), into
Eq.~(\ref{3corint}), we obtain \bea \label{sciI}
&&\chi(\omega)=(-1)^n \langle \vec{\sigma}_{n}
(\vec{\sigma}_{n+1}\times\vec{\sigma}_{n+2})\rangle\\
&=&\left[\frac {2}{\omega}I(0)
-\frac{2+4\omega^2}{\omega}I(\omega)
 -(1+\omega^2)\partial_\omega
I(\omega) \right].\nonumber \eea This function is
plotted in Fig. 2. We see a perfect match between our plot and the
numerical simulations of Ref.~\onlinecite{frahm}.

In conclusion, let us briefly comment on the possibility of
extension of the developed method to the third neighbor
correlation functions, in particular of  the type, $\langle
\vec{\sigma}_{n} \vec{\sigma}_{n+3} \rangle$. For this case one
has to increase the level of inhomogeneity of the model by
introducing two different shifts of the spectral parameter and
consider the following monodromy matrix \bea \label{3tm}
\Theta(\lambda;\omega_1,\omega_2)&=&
L_{3N,a}(\lambda)L_{3N-1,a}(\lambda-\omega_1) L_{3N-2,a}(\lambda-\omega_2)\nonumber \\
&\cdots& L_{3,a}(\lambda)L_{2,a}(\lambda-\omega_1)
L_{1,a}(\lambda-\omega_2), \eea which is defined on the lattice
with $3N$ sites. With this construction, one again has an
integrable model with commuting family of transfer matrices, but,
contrary to case considered above, we will have now three local
Hamiltonian operators. It is straightforward to derive their
explicit forms, which are rather cumbersome and we don't bring
them here. These operators contain different products of spins
residing on four neighboring sites, including, e.g., the
combination $ \vec{\sigma}_{n} \vec{\sigma}_{n+3}$. The number of
quasi- shift operators will be also three instead of the two as in
Eq.~(\ref{shops}). It is plausible to think that one can use
relations analogous to Eqs. (\ref{o1})- (\ref{do}) in order to
extract contributions of different summands out of the local
Hamiltonian operators, at least in the homogeneous limit,
$\omega_1=\omega_2=0$. These investigations may constitute
separate article.

\begin{figure}[t]
\centerline{\includegraphics[width=55mm,angle=0,clip]{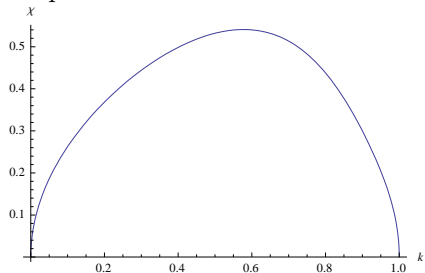}}
\caption{Chiral-spin order parameter, $\chi$, versus
$k=\frac{\omega^2}{1+\omega^2}.$}
\end{figure}

The authors acknowledge the discussions with  A. Nersesyan and T.
Sedrakyan with thanks. V.M. acknowledges ICTP and SISSA for
hospitality where the part of this work was done and INTAS grant
YS-05-109-5041.

\end{document}